\documentclass[10pt,english]{article}
\usepackage{subfigure}
\usepackage{graphicx}
\usepackage{float}
\usepackage[T1]{fontenc}
\usepackage[latin9]{inputenc}
\usepackage[margin=1.25in]{geometry}
\usepackage{float}
\usepackage{amsmath}
\usepackage{amsbsy}
\usepackage{setspace}
\usepackage{amssymb}
\usepackage{esint}
\usepackage{cite}
\newfloat{algorithm}{tbp}{loa}
\floatname{algorithm}{Algorithm}
\usepackage{algorithmic}

\newlength\myindent
\makeatletter

\floatstyle{ruled}
\newfloat{algorithm}{tbp}{loa}
\floatname{algorithm}{Algorithm}

\usepackage{babel}

\begin{document}

\title{On the inherent competition between valid and spurious\\ inductive inferences in Boolean data}

\author{M. Andrecut}

\date{August 15, 2017}

\maketitle
{

\centering Calgary, Alberta, T3G 5Y8, Canada

\centering mircea.andrecut@gmail.com

} 
\bigskip 
\begin{abstract}
Inductive inference is the process of extracting general rules from specific observations. 
This problem also arises in the analysis of biological networks, such as genetic regulatory networks, where 
the interactions are complex and the observations are incomplete. 
A typical task in these problems is to extract general interaction rules as combinations of Boolean covariates, 
that explain a measured response variable. 
The inductive inference process can be considered as an incompletely specified Boolean function synthesis problem. 
This incompleteness of the problem will also generate spurious inferences, 
which are a serious threat to valid inductive inference rules.  
Using random Boolean data as a null model, here we attempt to measure the competition between valid and spurious inductive inference rules from a given data set. 
We formulate two greedy search algorithms, which synthesize a given Boolean response variable in a sparse 
disjunct normal form, and respectively a sparse generalized algebraic normal form of the variables from the observation data, 
and we evaluate numerically their performance. 
\smallskip


\smallskip

\end{abstract}

\bigskip

\section{Introduction}

Inductive inference is a frequently used method in the analysis of biological networks (such as genetic regulatory networks),  
or for explaining drug responses to logic combinations of perturbations.\cite{key-1}  
In many such analysis problems the predictors and the targets (genes, perturbations, drug responses) are modeled as Boolean variables (active-inactive, present-absent), 
and they are characterized by complex non-linear interactions. These problems are also typically ill conditioned, since the number of observations 
is much smaller than the number of possible logic rules describing the interactions between the variables. 

Contrary to the deductive inferences, which are true if their premises are true (a complete set of observations is required in this case), 
the inductive inferences are inherently uncertain. 
Thus, the logical inductive rules extracted from such complex interacting networks are only probable, given the incomplete data observations. 
Nevertheless, the extraction of such inductive inference rules is important for logic interaction hypothesis formulation, and for network reverse-engineering. Relevant examples include identification of: 
regulatory motifs,\cite{key-2} 
drug active pathways from gene networks estimated by gene expression data,\cite{key-3}  
association of gene-gene interactions,\cite{key-4}  
single nucleotide polymorphism (SNPs) in genome-wide association studies,\cite{key-5,key-6} 
combinatorial effects in cancer biology,\cite{key-7,key-8}  
dynamic biological networks based on responses to drug perturbations.\cite{key-9}

To better understand these problems, let us consider the typical example of inferring logical rules explaining 
a response variable as a function of logic combination of perturbations. In such a problem we have a $M\times N$ Boolean matrix $X=[x_{mn}]$, 
containing the $N$ logic perturbations in $M$ specific samples (observations), and a Boolean vector $y$ of length $M$ corresponding to the measured response. 
The task is to infer the simplest logical rule that explains $y$ given $X$. 
Such a problem is ill conditioned because the number of specific samples is typically much smaller than the number of 
all the possible logic interactions among the variables: $M\ll 2^N$. 

The ill conditioning and incompleteness of the problem will also generate spurious inferences, 
which are a serious threat to valid inductive inference rules.  
In fact for $N$ Boolean variables one can define 
$2^{2^N}$ Boolean functions. Among these functions there are $2^{2^N-M}$ functions which will satisfy the $M$ constraints (samples or observations). 
Thus, the probability of "guessing" such a "spurious" function is $2^{-M}$, and exponentially decreases with the number of constraints. 
Using random Boolean data as a null model, here we attempt to measure the competition between valid and spurious inductive inference rules from a given data set. 
We formulate two greedy search algorithms, which synthesize a given Boolean response variable in a sparse 
Disjunct Normal Form (DNF)\cite{key-10}, and respectively a sparse 
Generalized Algebraic Normal Form (GANF)\cite{key-11}  of the variables from the observation data, 
and we evaluate numerically their performance. 

\section{Inductive inference as a function synthesis problem}

The inductive inference process can be considered as an incompletely specified Boolean Function Synthesis (BFS) problem.\cite{key-10} 
In this case, given the data $X$ and the response vector $y$, the task is to find the simplest Boolean function $f:\{0,1\}^N \rightarrow {0,1}$, 
such that:
\begin{equation}
f(x_{m1},x_{m2},...,x_{mN})=y_m,\quad m=1,2,...,M.
\end{equation}
Obviously such a function is incompletely specified by the matrix $X$, 
since $M\ll 2^N$, where $2^N$ is the number of total possible combinations between the $N$ Boolean variables. 
Using Boolean algebra it can be shown that any Boolean function $f$ can be represented using the Canonical DNF (CDNF) expansion.\cite{key-10} However, 
in the CDNF each disjunct clause contains all $N$ variables, and the number of such disjuncts is equal to the number of observations when $f$ is 1 (TRUE).\cite{key-10} 
Obviously such a large expansion does not provide a practical interpretation, and therefore we are interested in finding the DNF expansion with the smallest 
number of the smallest disjuncts, which from the bio-medical complexity point of view gives a more appropriate interpretation (parsimony principle). 
This is a hard NP-complete problem, and several heuristic algorithms have been formulated to solve it approximatively.\cite{key-10} 
An important relaxation of this problem is to find the simplest K-DNF, which is a disjunction of conjuctive clauses (sum of products) 
containing at most $K$ variables, where each variable may appear as either complemented or not complemented:
\begin{align}
f_{DNF}(x_1,x_2,...,x_N) &= a_1 \dot{x}_1 + a_1 \dot{x}_2 + ... + a_N \dot{x}_N + \nonumber \\
 				   &\quad\; a_{1,2} \dot{x}_1 \dot{x}_2 + ... + a_{N-1,N} \dot{x}_{N-1}\dot{x}_{N} + \nonumber \\
 				   &\quad\;...+ a_{i_1,i_2,...,i_K} \dot{x}_{i_1} \dot{x}_{i_2} ... \dot{x}_{i_K}
\end{align}
where $a_1,a_2,...,a_{i_1,i_2,...,i_K} \in \{0,1\}$, $\dot{x}_n \in \{x_n,\bar{x}_n\}$, $\bar{x}_n$ is the complement of $x_n$, and $+$ is the OR operator. 

It can be shown also that any Boolean function $f$ of $N$ variables can be represented using the GANF expansion, 
which is a modulo-2 (XOR $\oplus$) sum of products.\cite{key-11} 
Similarly to the DNF expansion, we can write the K-GANF expansion:
\begin{align}
f_{GANF}(x_1,x_2,...,x_N) &= b_1 \dot{x}_1 \oplus b_1 \dot{x}_2 \oplus ... \oplus b_N \dot{x}_N \oplus \nonumber \\
 				   &\quad\; b_{1,2} \dot{x}_1 \dot{x}_2 \oplus ... \oplus b_{N-1,N} \dot{x}_{N-1}\dot{x}_{N} \oplus \nonumber \\
 				   &\quad\;...\oplus b_{i_1,i_2,...,i_K} \dot{x}_{i_1} \dot{x}_{i_2} ... \dot{x}_{i_K}
\end{align}
where $b_1,d_2,...,b_{i_1,i_2,...,i_K} \in \{0,1\}$ and $\dot{x}_n \in \{x_n,\bar{x}_n\}$.
One can see that the GANF expansion is fundamentally different than DNF, because of the injectivity of the XOR operation. 

We should note that GANF is also known in the literature as the Generalized Reed-Muller Normal Form (GRMNF)\cite{key-12,key-13},  
and it has been shown that for a small number of variables the minimal GRMNF expansion can be calculated efficiently using the exact algorithm 
given in Ref.\cite{key-12}, 
or the Walsh transform approach, given in Ref.\cite{key-13}. 
However, for a larger number of variables there is no straightforward algorithm for finding the minimal GANF (GRMNF)
except, for an exhaustive search.

Here, we formulate two greedy search algorithms that attempt to find sparse K-DNF and K-GANF expansions of a given response function, and therefore to 
extract simple inductive inference rules as combinations of Boolean covariates from incomplete observations. 
Thus, the goal is to find the K-DNF and K-GANF expansions that best approximate an incomplete Boolean function $f$, and have the smallest number of TRUE coefficients 
$a_1,a_2,...,a_{i_1,i_2,...,i_K}$, and respectively $b_1,b_2,...,b_{i_1,i_2,...,i_K}$. 
In order to evaluate the performance of the algorithms,   
we use Boolean data generated by random Bernoulli variables, following the distribution:
\begin{equation}
x = Ber(p) \; \Leftrightarrow \; 
  \begin{cases}
   P(x=1) = p \\
   P(x=0) = 1-p
  \end{cases},\; 0<p<1
\end{equation} 

\section{Greedy search for inference rules}

In order to formulate the greedy algorithms we consider the set of all conjunctive clauses containing at most $K$ variables. 
The total number of such clauses grows very fast with $K$. Therefore, in order to limit their number, and to maintain the 
required simplicity for the solutions we consider only the cases with $K=1,2,3$. 

Given the observation data $X\in \{0,1\}^{M \times N}$ and the response $y\in \{0,1\}^{M}$, 
we expand the matrix $[X;\bar{X}]$ into the larger matrix $G\in \{0,1\}^{M \times Q}$ of all conjunctive clauses $g_q$, $q=1,2,...,Q$, containing at most $K$ variables, calculated over all $M$ observations from the matrix $[X;\bar{X}]$. 
It is also natural to exclude from $G$ all the conjunctive clauses that evaluate to 0 (FALSE) for all the observations,  
since obviously they have no contribution to the DNF and GANF expansions.

\subsection{Greedy K-DNF}

The algorithm is seeking a sparse K-DNF expansion of the form:
\begin{equation}
f = g_{q_1} + g_{q_2} + ... + g_{q_{j}}  = \sum_{i=1}^j a_{q_i}g_{q_i},
\end{equation}
where $q_i$ is the index of the selected clause $g_{q_i}$ from the matrix $G$ of size $M \times Q$, and $a_{q_i}$=TRUE, $i=1,2,...,j$. 
The goal of the algorithm is to minimize the Hamming distance between the K-DNF expansion $f \in \{0,1\}^M$ and the Boolean response $y \in \{0,1\}^M$:
\begin{equation}
\min \left\lbrace  d_H(y,f)= \sum_{m=1}^M y_m \oplus f_m \right\rbrace .
\end{equation}
The algorithm starts with completely FALSE (0) vectors $f$ ($f_m=0,m=1,2,...,M$) and $a$ ($a_q=0,q=1,2,...,Q$), and at each new step 
selects a new column (clause) $g_{q_j}$ from the matrix $G$, that minimizes the Hamming distance $d_H(y,f_{q_j})$:
\begin{equation}
q_j = \text{arg} \min d_H(y,f_{q_j}),
\end{equation}
where 
\begin{equation}
f_{q_j} = f_{q_{j-1}} + g_{q_j} = g_{q_1} + g_{q_2} + ... + g_{q_{j-1}} + g_{q_j}.
\end{equation}
The algorithm stops when $d_H(y,f_{q_j}) = 0$, or when no column $g_{q_j}$ could be found to minimize the Hamming distance. 
Thus, the resulted inductive inference rule is completely specified by the set of indexes 
$\{q_1,q_2,...,q_j\}$ of the clauses extracted from the matrix $G$. 
One can see that the approximation $f$ can be written using the linear matrix-vector multiplication $f = Ga$, 
where $a$ is a sparse vector, where the coefficients $\{a_{q_1},a_{q_2},...,a_{q_j}\}$ corresponding to the selected clauses are TRUE, while the rest are FALSE, 
and the sum $+$ corresponds to the the OR operator.

A detailed description of the method is given in Algorithm 1. 
The function KDNF() takes as input two bit arrays, 
one containing the clauses $g_q \in \{0,1\}^M$, $q=1,2,...,Q$, and the other one containing the response variable $y \in \{0,1\}^M$, over the $M$ different measurements. 
In the next step the algorithm initializes the variables $a,f,D_{min},j$. 
Here, $a$ is a bit array corresponding to the coefficients in the K-DNF expansion, $f=\text{falses}(M)$ is the initial approximation, 
$D_{min}=M$ is the maximum Hamming distance between $f$ and $y$, and $j=0$ is the initial number of clauses included in the expansion. 
The algorithm continues with a loop where maximum $M$ clauses can be selected for the K-DNF expansion. At each step the column $q$ that minimizes 
the Hamming distance $d=d_H(f + g_q,y)$ is selected, and $D$ is updated with the minimum value $d$. 
If $D<D_{min}$ then $j$ is increased with one unit, 
the variables are updated $a_q$, $f$, $D_{min}$ are updated with TRUE, $f + g_q$, and respectively $D$. 
The function returns: $f$, the approximation of the response variable $y$; $a$, the coefficients of the K-DNF expansion;  
$D_{min}$ the Hamming distance between $f$ and $y$; and $j$, the number of clauses included in the K-DNF expansion of $y$. 

\begin{algorithm}[t!]
\caption{Greedy search for sparse K-DNF expansion.}
\begin{algorithmic}[1]
\STATE $\textbf{function} \text{ KDNF}(g::BitArray,y::BitArray)$
\STATE $M,Q \leftarrow$ size$(g)$
\STATE $a,f \leftarrow$ falses$(Q)$,falses$(M)$
\STATE $D_{min},j \leftarrow M,0$
\FOR{$t=1:M$}
	\STATE $q,D \leftarrow 0,D_{min}$
		\FOR{$n=1:Q$}
			\STATE $d \leftarrow$ $d_H(f + g_n,y)$
			\IF{$d < D$}
				\STATE $q,D \leftarrow n,d$
			\ENDIF
		\ENDFOR
	\IF{$D < D_{min}$}
		\STATE $j \leftarrow j + 1$
		\STATE $a_q \leftarrow $TRUE
		\STATE $f \leftarrow f + g_q$
		\STATE $D_{min} \leftarrow D$
	\ELSE
		\STATE break
	\ENDIF
\ENDFOR
\RETURN $f, a, D_{min}, j$
\STATE $\textbf{end function}$
\end{algorithmic}
\end{algorithm}

\bigskip

\subsection{Greedy K-GANF}

Let us assume that $f$ is the approximation of $y$, such that the residual error is:
\begin{equation}
r = y \oplus f.  
\end{equation} 
Reciprocally we also have:
\begin{equation}
f = r \oplus y.
\end{equation} 
Also, we can write the approximate solution as following:
\begin{equation}
f_{q_j} = \bigoplus_{i=1}^j b_{q_i}g_{q_i} = g_{q_1} \oplus g_{q_2} \oplus ... \oplus g_{q_{j}} = f_{q_{j-1}} \oplus g_{q_{j}},
\end{equation}
where $q_{q_j}$ are the clauses (columns) selected from the matrix $G$, $b_{q_i}=$TRUE, $i=1,2,...,j$.
The algorithm seeks to minimize the norm of the residual:
\begin{equation}
r_{q_j} = y \oplus f_{q_j} = y \oplus f_{q_{j-1}} \oplus g_{q_{j}} = r_{q_{j-1}} \oplus g_{q_{j}}.
\end{equation}
Thus, the algorithm starts with $r = y$ and a FALSE (0) vector $b$ ($b_q=0,q=1,2,...,Q$), and at each new step selects the new column (clause) $g_{q_j}$ that minimizes the 
norm of the new residual $r$:
\begin{equation}
q_j = \text{arg} \min |r|,
\end{equation}
where 
\begin{equation}
r \leftarrow r \oplus g_{q_j}, 
\end{equation}
and 
\begin{equation}
|r| = \sum_{m=1}^M r_m.
\end{equation}
The algorithm stops when $|r| = 0$, or when no column $g_{q_j}$ could be found to minimize the norm. The final approximation of the Boolean response is:
$f = r \oplus y$.
Again, the resulted inductive inference rule is completely specified by the set of indexes 
$\{q_1,q_2,...,q_j\}$ of the clauses selected from the matrix $G$, such that the corresponding K-GANF expansion coefficients $\{b_{q_1},b_{q_2},...,b_{q_j}\}$ are TRUE, 
while the rest are FALSE, such that we have $f=Gb$, where the sum $+$ is replaced by the $\oplus$ operator.

A detailed description of the method is given in Algorithm 2. 
The function KGANF() takes as input two bit arrays, 
one containing the clauses $g_q \in \{0,1\}^M$, $q=1,2,...,Q$, and the other one containing the response variable $y \in \{0,1\}^M$, over the $M$ different measurements. 
In the next step the algorithm initializes the variables $b,r,D,j$. 
Here, $b$ is a bit array corresponding to the coefficients in the K-GANF expansion, $r=y$ is the initial residual, 
$D_{min}=M$ is set to the maximum possible norm of $r$, and $j=0$ is the initial number of clauses included in the expansion. 
The algorithm continues with a loop where maximum $M$ clauses can be selected for the K-GANF expansion. At each step the column $q$ that minimizes 
the norm of $r \oplus g_q$ is selected, and $D_{min}$ is updated with the minimum value $d$. 
If the column $q$ was already selected in a previous step ($b_q=$TRUE) then $j$ is decreased, since the XOR operator $\oplus$ applied a second time  
with the same data cancels the previous application. If the column $q$ was not selected before ($b_q=$FALSE) then $j$ is increased with one unit. 
The values of $b_q$ and $r$ are updated with $b_q \oplus $TRUE and respectively $r \oplus g_q$. 
The algorithm stops if the norm of the residual becomes zero, $D_{min}=|r|=0$, or no other column to minimize $|r|$ could be found, $q=0$. 
The function returns: $f =r \oplus y$, the approximation of the response variable $y$; $b$, the coefficients of the K-GANF expansion;  
$D_{min}$ the norm of the residual $r$; and $j$, the number of clauses included in the K-GANF expansion of $y$. 

\begin{algorithm}[t!]
\caption{Greedy search for sparse K-GANF expansion.}
\begin{algorithmic}[1]
\STATE $\textbf{function} \text{ KGANF}(g::BitArray,y::BitArray)$
\STATE $M,Q \leftarrow$ size$(g)$
\STATE $b,r,D_{min},j \leftarrow$ falses$(Q)$,$y,M,0$
\FOR{$t=1:M$}
	\STATE $q \leftarrow 0$
		\FOR{$n=1:Q$}
			\STATE $d \leftarrow$ $|r \oplus g_n|$
			\IF{$d\leq D_{min}$}
				\STATE $q,D_{min} \leftarrow n,d$
			\ENDIF
		\ENDFOR
	\IF{$q > 0$}
		\IF{$b_q$}
			\STATE $j \leftarrow j - 1$
		\ELSE
			\STATE $j \leftarrow j + 1$
		\ENDIF
		\STATE $b_q \leftarrow b_q \oplus $TRUE
		\STATE $r \leftarrow r \oplus g_q$
	\ENDIF
	\IF{$D_{min} = 0$ or $q = 0$}
		\STATE break
	\ENDIF
\ENDFOR
\RETURN $r \oplus y, b, D_{min}, j$
\STATE $\textbf{end function}$
\end{algorithmic}
\end{algorithm}
\bigskip

\begin{figure}[t!]
  \centering
  \subfigure{\includegraphics[scale=0.55]{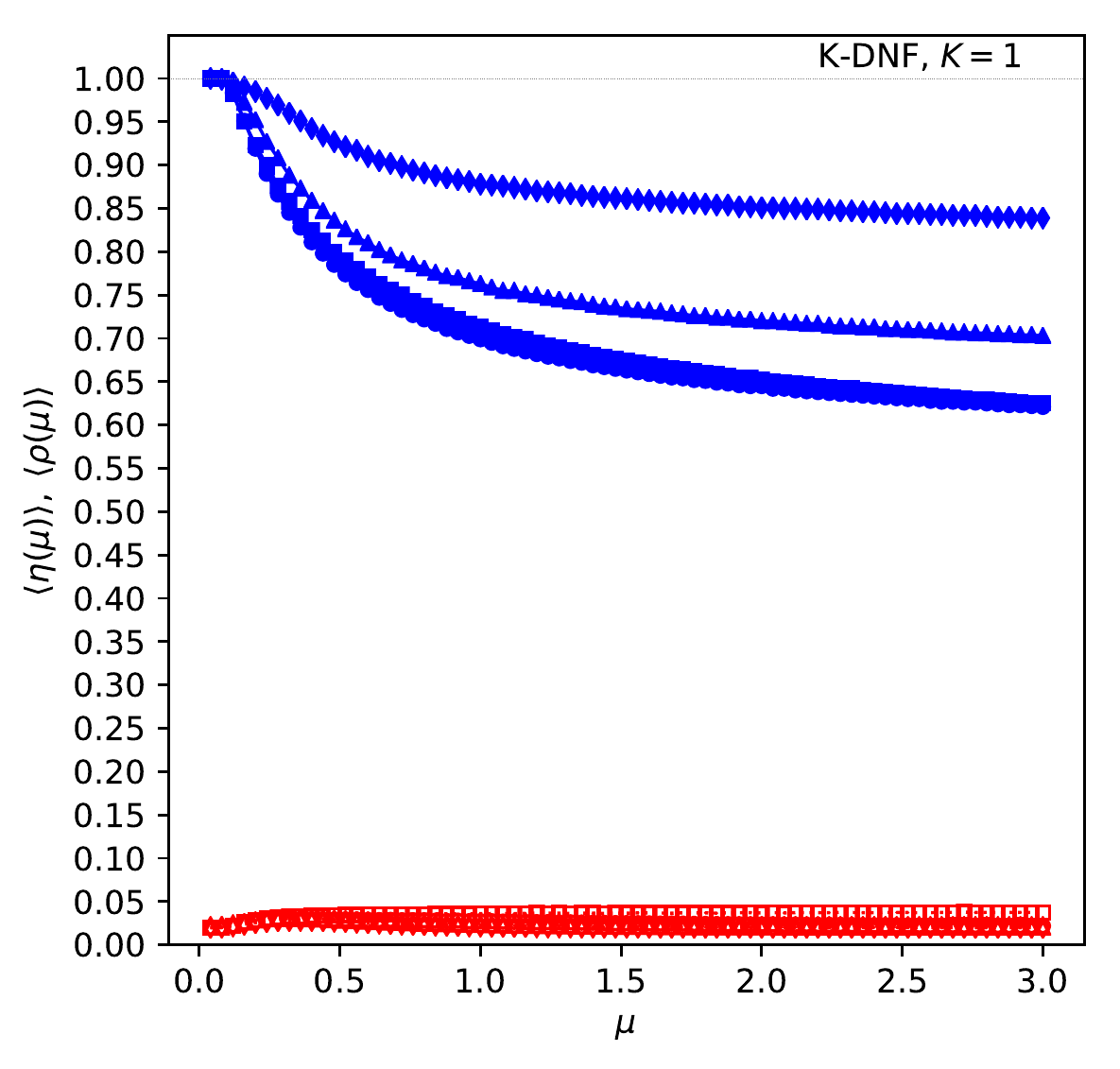}}
  \subfigure{\includegraphics[scale=0.55]{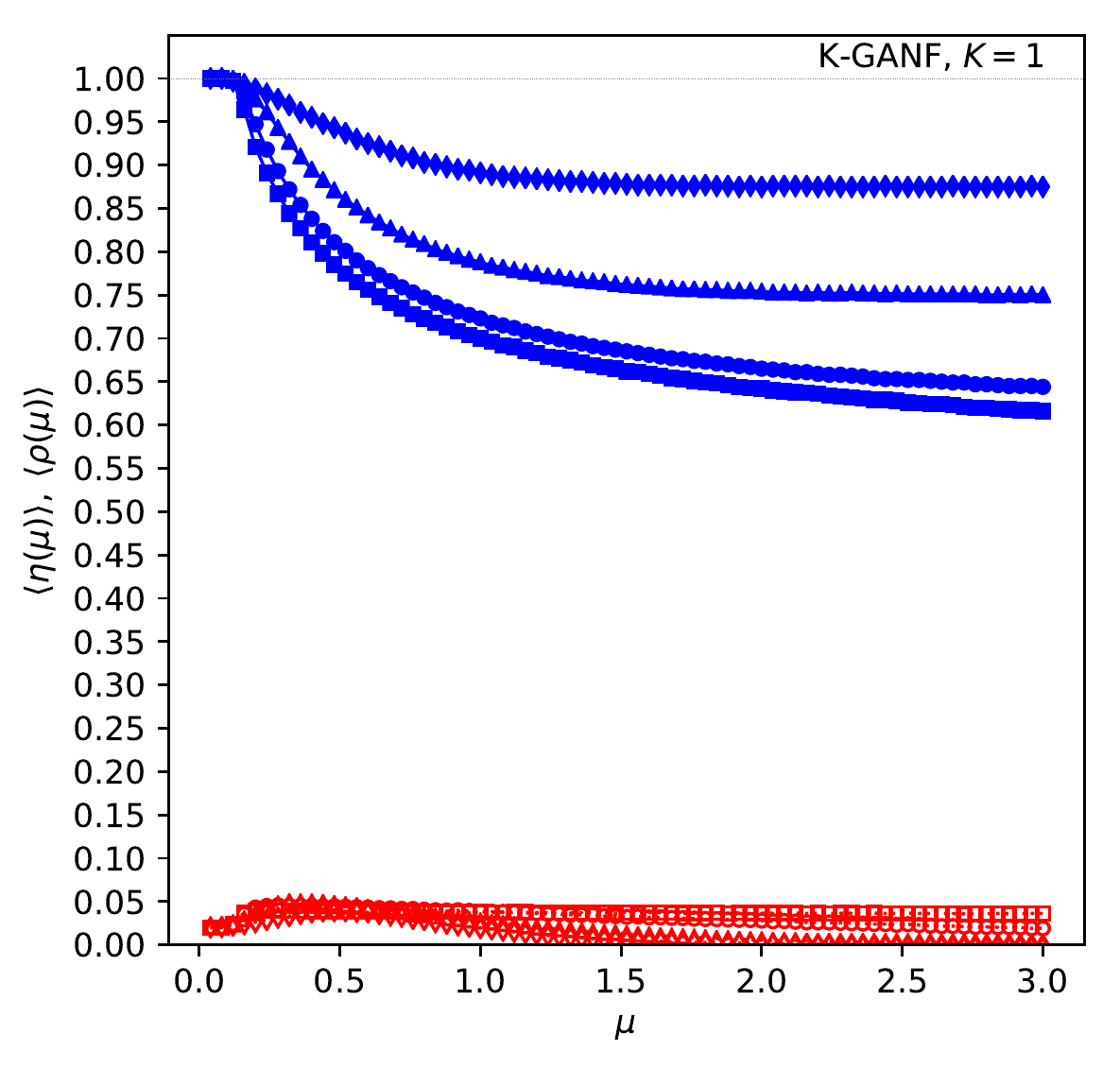}}\\[-2.5ex]
  \subfigure{\includegraphics[scale=0.55]{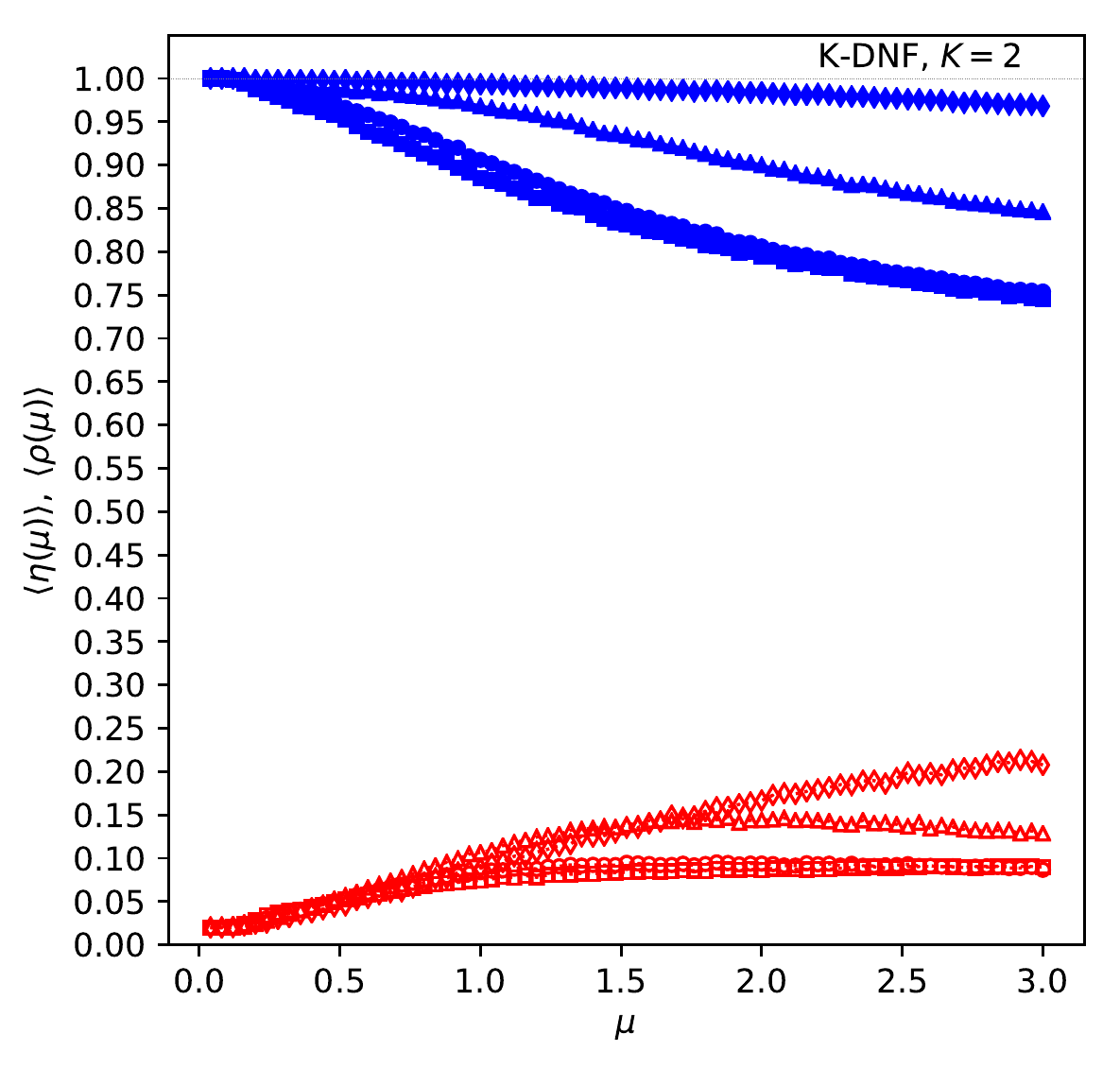}}
  \subfigure{\includegraphics[scale=0.55]{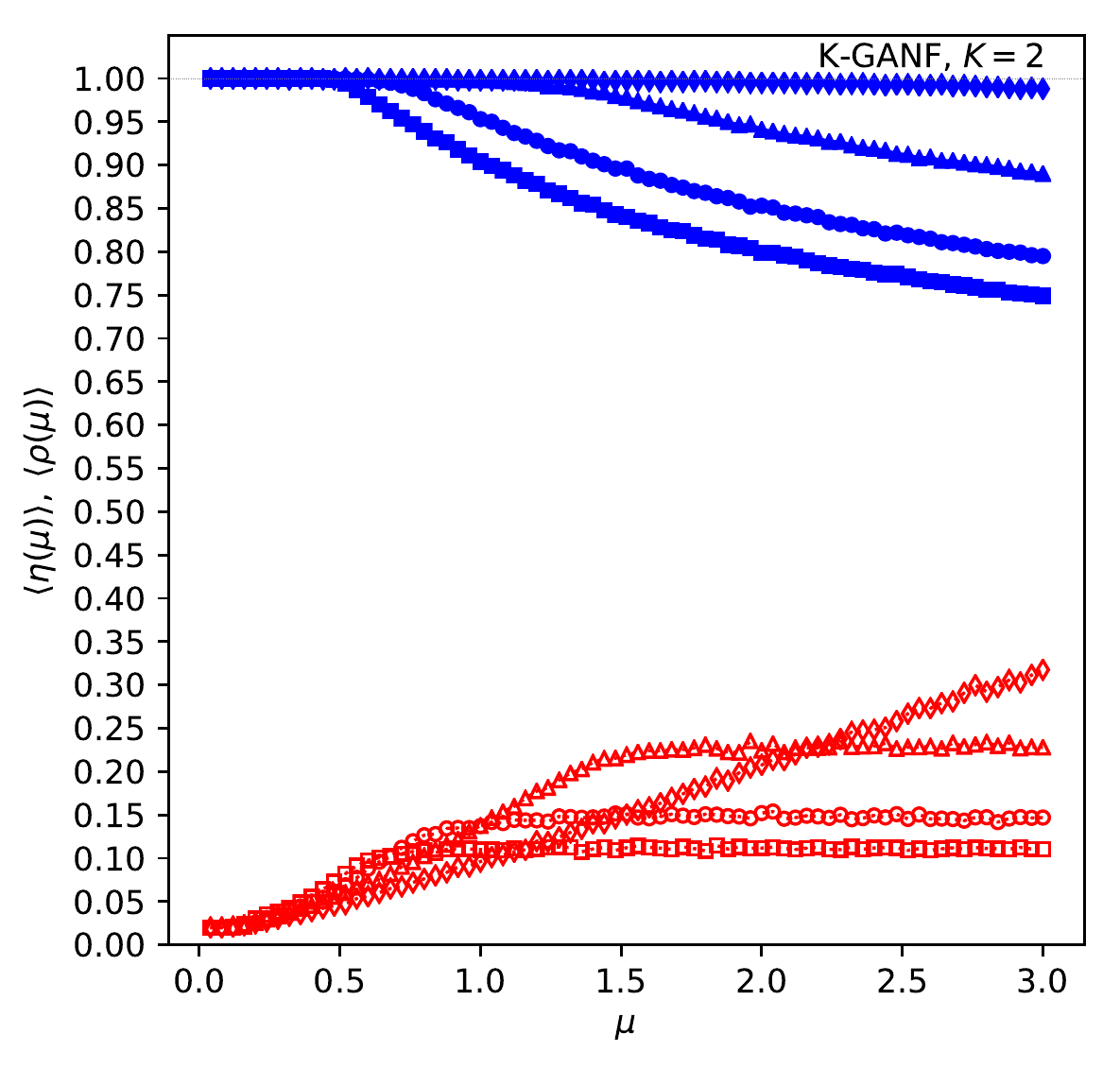}}\\[-2.5ex]
  \subfigure{\includegraphics[scale=0.55]{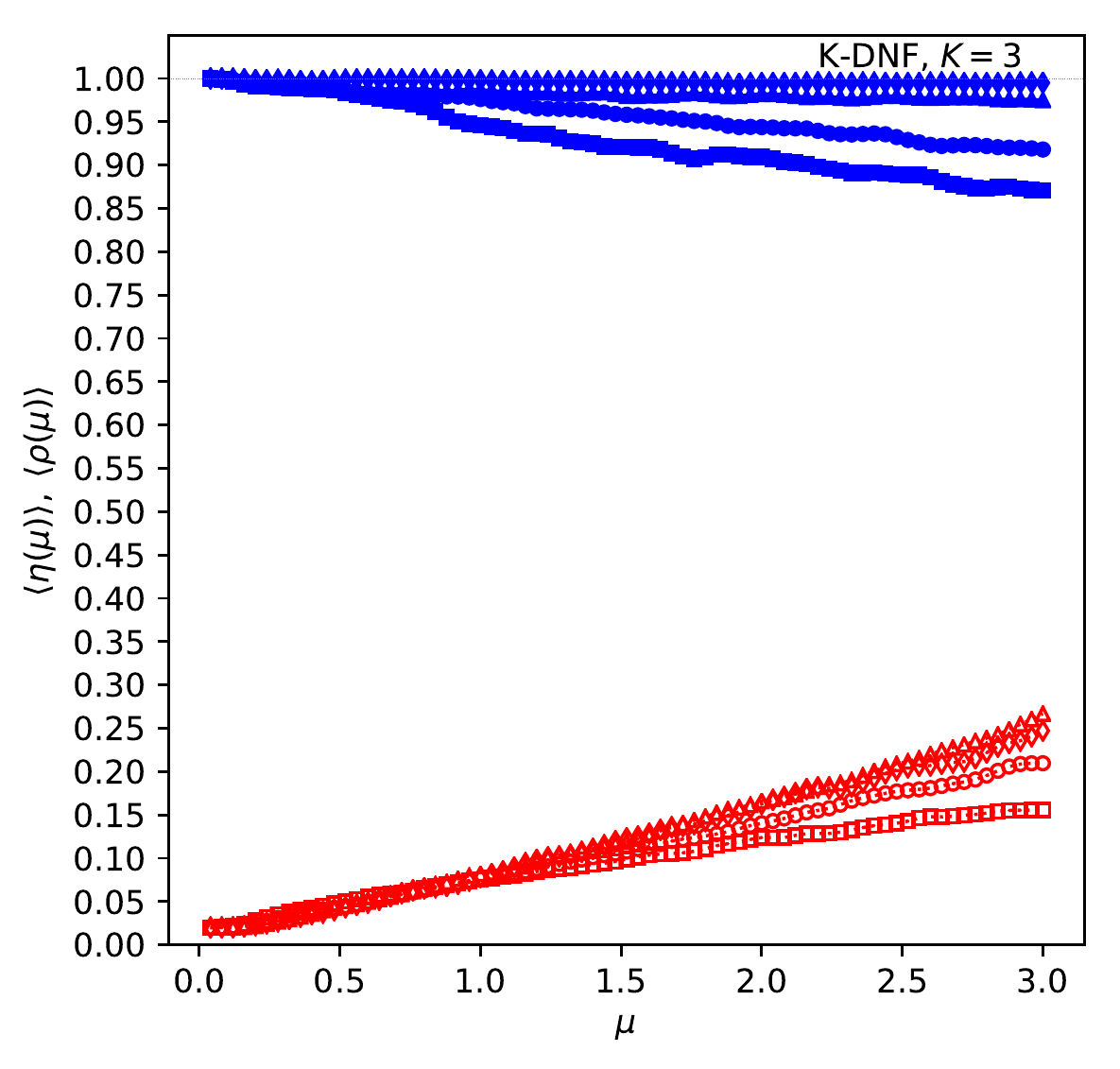}}
  \subfigure{\includegraphics[scale=0.55]{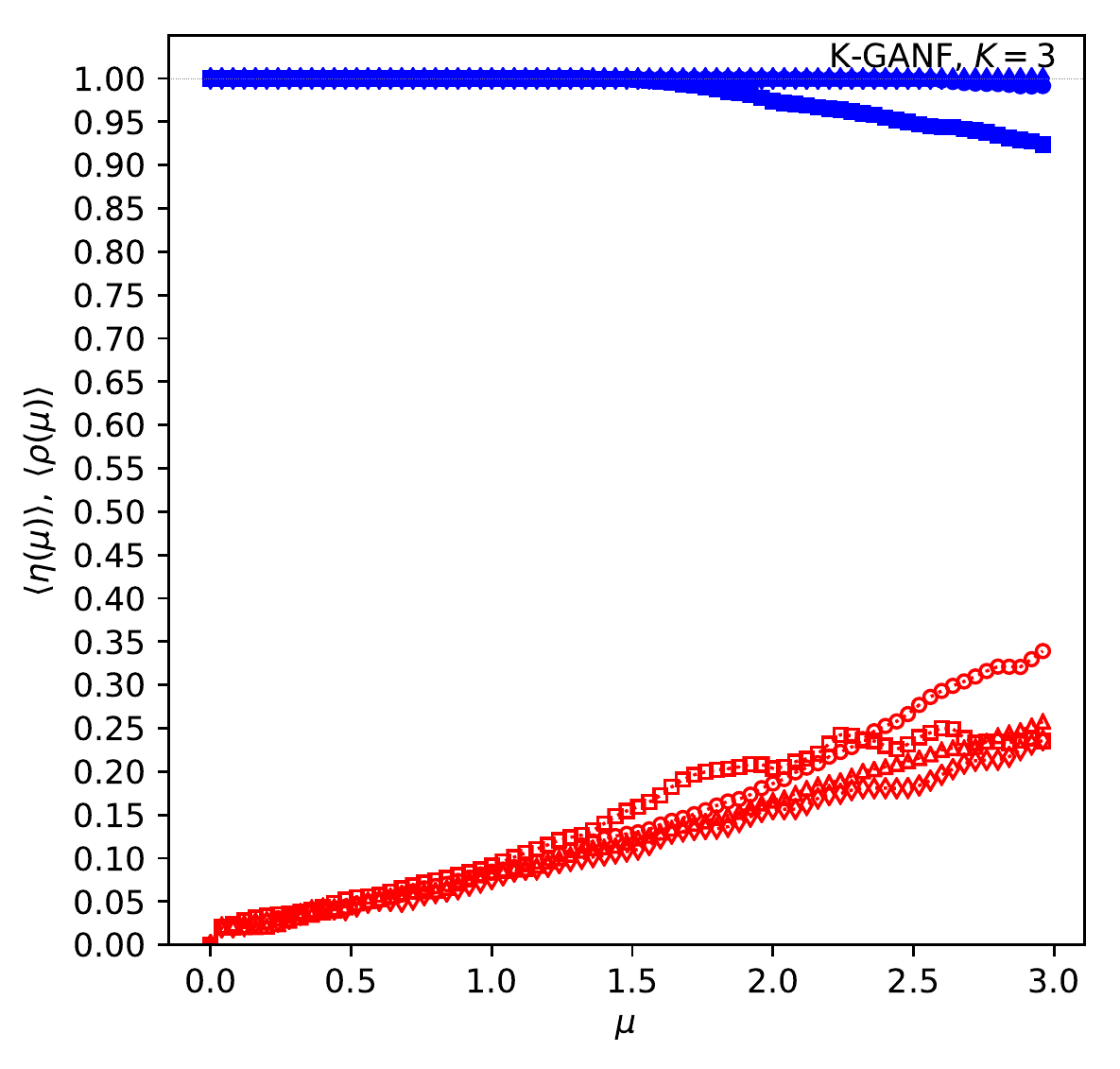}}
\caption{The probability  of exact synthesis $\langle \eta(\mu) \rangle$ (blue filled symbols) and 
the average number of clauses $\langle \rho(\mu) \rangle$ (red empty symbols) as a function of 
$\mu=M/N$ and the probability $p$ of the Bernoulli processes ($p=0.5$ squares, $p=0.375$ circles, $p=0.25$ triangles, $p=0.125$ diamonds). 
K-DNF left column, K-GANF right column, $K=1$ top, $K=2$ middle, $K=3$ bottom. 
}
\end{figure}

\section{Numerical results}
\subsection{Sparse Boolean function synthesis}
In order to illustrate numerically the performance of the GS algorithm we consider $N=50$ Boolean variables. 
We simulate the problem for Bernoulli processes with $0<p\leq 0.5$, using $10^3$ different instances for each $\mu = M/N \in [0,3]$. 

During the simulation we collect the probability of exact synthesis, $\eta = 1 - d_H(f,y)/M$, 
where $d_H(f,y)$ is the Hamming distance between the best approximation $f$ and the target $y$, while $M$ is the maximum possible Hamming distance. 
We also collected $\rho = j/N$, where $j$ the number of required clauses, and $N$ is the number of variables in the matrix $X$.
Finally we represent graphically the average quantities $\langle \eta(\mu) \rangle$ 
and respectively $\langle \rho(\mu) \rangle$, as a function of $\mu = M/N$. 

The simulation results are shown in Figure 1.
One can see that both algorithms can synthesize very sparse expansions with high probabilities. 
The synthesis probability $\langle \eta(\mu) \rangle$ exhibits a transition as a 
function of $\mu$ and $K$, and the transition point also depends on the probability $p$ of the Bernoulli process used to generate the Boolean data, $\mu^*(p)$. 
As expected, the synthesis probability increases as $p$ decreases. 

\pagebreak

For $K=1$ the synthesis probability deteriorates relatively fast for $\mu^*>0.2$. 
For $K=2$ the number of possible choices for the clauses in the expansion is quadratically larger ($\sim N^2$), and this 
reflects also on higher values of the transition point $\mu^*$. 
The results are more dramatic for $K=3$, since for this case the number of possible choices for the clauses is much larger ($\sim N^3$), 
resulting in much larger values of the transition point $\mu^*$.
The transition point $\mu^*$ corresponds to a synthesis probability of $\eta(\mu^*)=0.98$, which also corresponds to a 1 bit synthesis error in the response function $y$. 
The obtained values of the transition points $\mu^*$ are summarized in Table 1, and respectively in Table 2. 

\begin{table}[h]
\caption{K-DNF transition points $\mu^*$ for 1-bit synthesis error $\eta(\mu^*)=0.98$.}
\smallskip
\centering
{
\begin{tabular}{l*{10}{c}r}
$K$ & $p=0.125$ & $p=0.250$ & $p=0.375$ & $p=0.500$ \\
\hline
1 & 0.24 & 0.16 & 0.14 & 0.12 \\ 
2 & 2.36 & 0.80 & 0.42 & 0.29 \\ 
3 & $>3.0$ & 1.54 & 1.01 & 0.41 \\ 
\end{tabular}
}
\end{table}
\begin{table}[!h]
\caption{K-GANF transition points $\mu^*$ for 1-bit synthesis error $\eta(\mu^*)=0.98$.}
\smallskip
\centering
{
\begin{tabular}{l*{10}{c}r}
$K$ & $p=0.125$ & $p=0.250$ & $p=0.375$ & $p=0.500$ \\
\hline
1 & 0.28 & 0.20 & 0.16 & 0.14 \\ 
2 & $>3.0$ & 1.50 & 0.84 & 0.60 \\ 
3 & $>3.0$ & $>3.0$ & 2.72 & 1.59 \\ 
\end{tabular}
}
\end{table}

\subsection{Sparse Boolean function recovery}

In a second numerical experiment we consider the recovery problem. More exactly, given the matrix 
$G\in \{0,1\}^{M \times Q}$ of all conjunctive clauses $g_q$, $q=1,2,...,Q$, we synthesize the response function $y$ 
using a sparse selection of $S$ clauses randomly drawn (with equal probability) from $g_q$, $q=1,2,...,Q$. 
Therefore, in this case the response $y$ is a priori synthesized using clauses drawn from $G$, and 
the task is to recover these unknown clauses using the K-DNF and K-GANF algorithms. 
Here we consider the cases for $S=1,2,3,4$ unknown clauses, with $N=50$ and $p=0.5$. As before, we average over $10^3$ different 
instances for each $\mu = M/N \in [0,3]$. We collect both the probability of exact synthesis $\langle \eta(\mu) \rangle$ 
and the probability of exact recovery $\langle \sigma(\mu) \rangle$. Here, $\sigma(\mu)$ measures the probability that all the 
unknown clauses in $y$ are correctly recovered by the K-DNF and K-GANF algorithms. 

The results are shown in Figure 2.
One can see that while the synthesis probability $\langle \eta(\mu) \rangle$ is very high for all $S$ values, 
the probability of exact recovery $\langle \sigma(\mu) \rangle$ substantially drops by increasing the number $S$ of unknown clauses. 
This means that the algorithms are able to synthesize a relatively good response, that matches well the given $y$, but 
they don't include the correct clauses. This is a consequence of the fact that the matrix $G$ may contain several "equivalent" clauses, 
competing in the greedy selection mechanism used by the algorithms, and which therefore lead to "spurious" inferences. 
Also, the correct recovery probability $\langle \sigma(\mu) \rangle$ drops sharply at small $\mu$ values, which is expected since the number of 
constraints $M$ is small for a small $\mu$, and therefore the probability of "spurious" inferences increases. 

\begin{figure}[t!]
  \centering
  \subfigure{\includegraphics[scale=0.55]{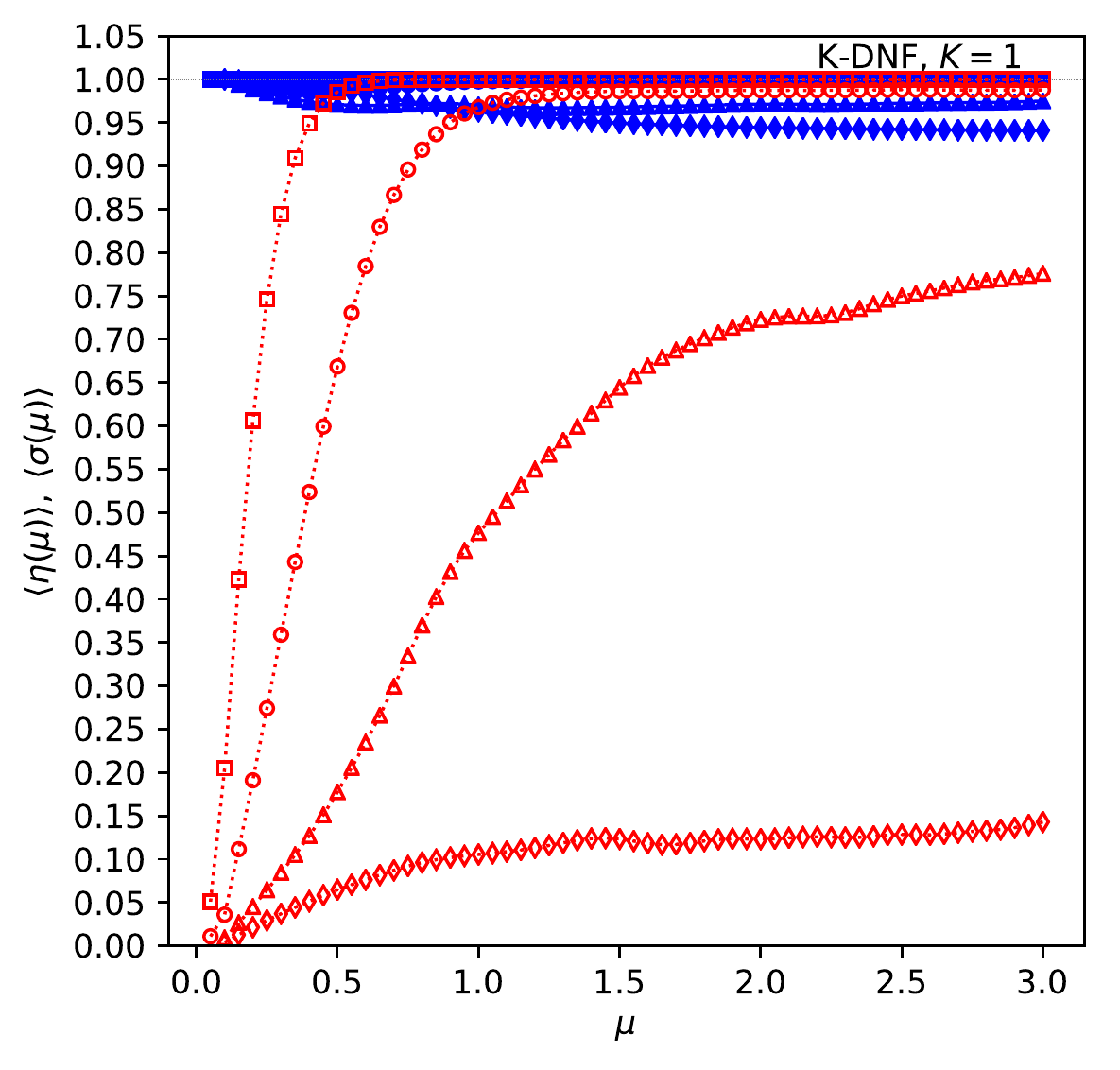}}
  \subfigure{\includegraphics[scale=0.55]{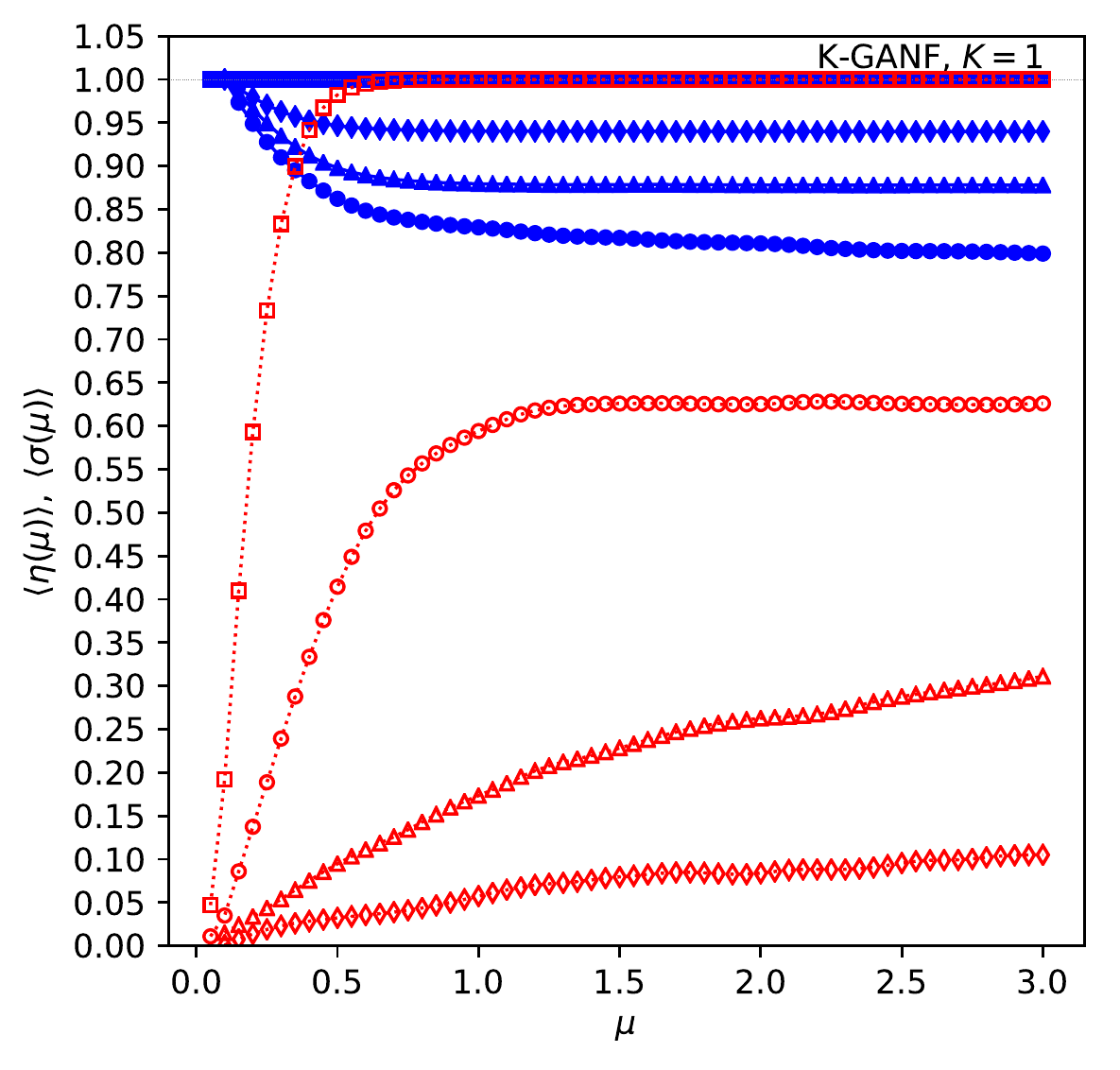}}\\[-2.5ex]
  \subfigure{\includegraphics[scale=0.55]{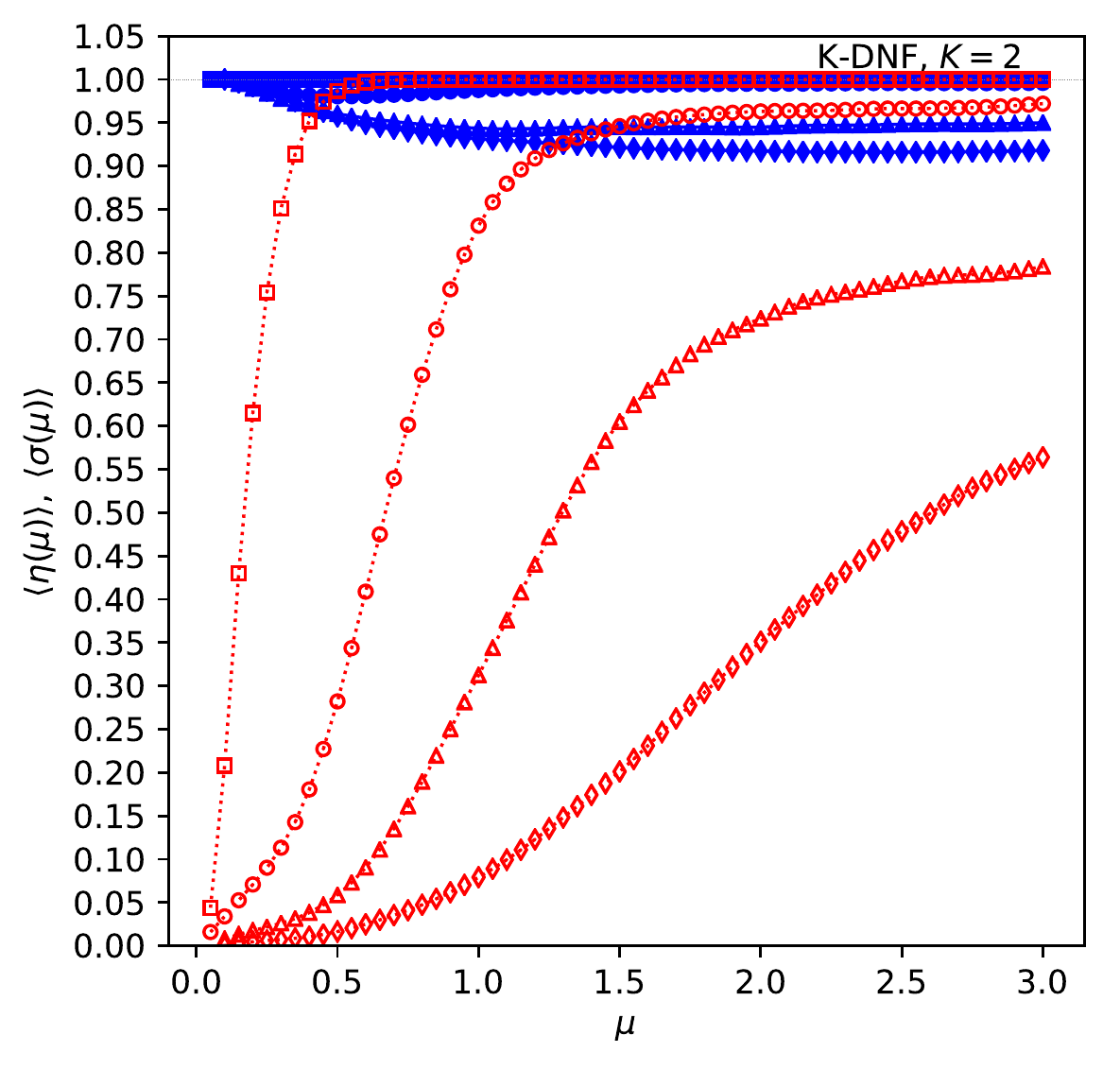}}
  \subfigure{\includegraphics[scale=0.55]{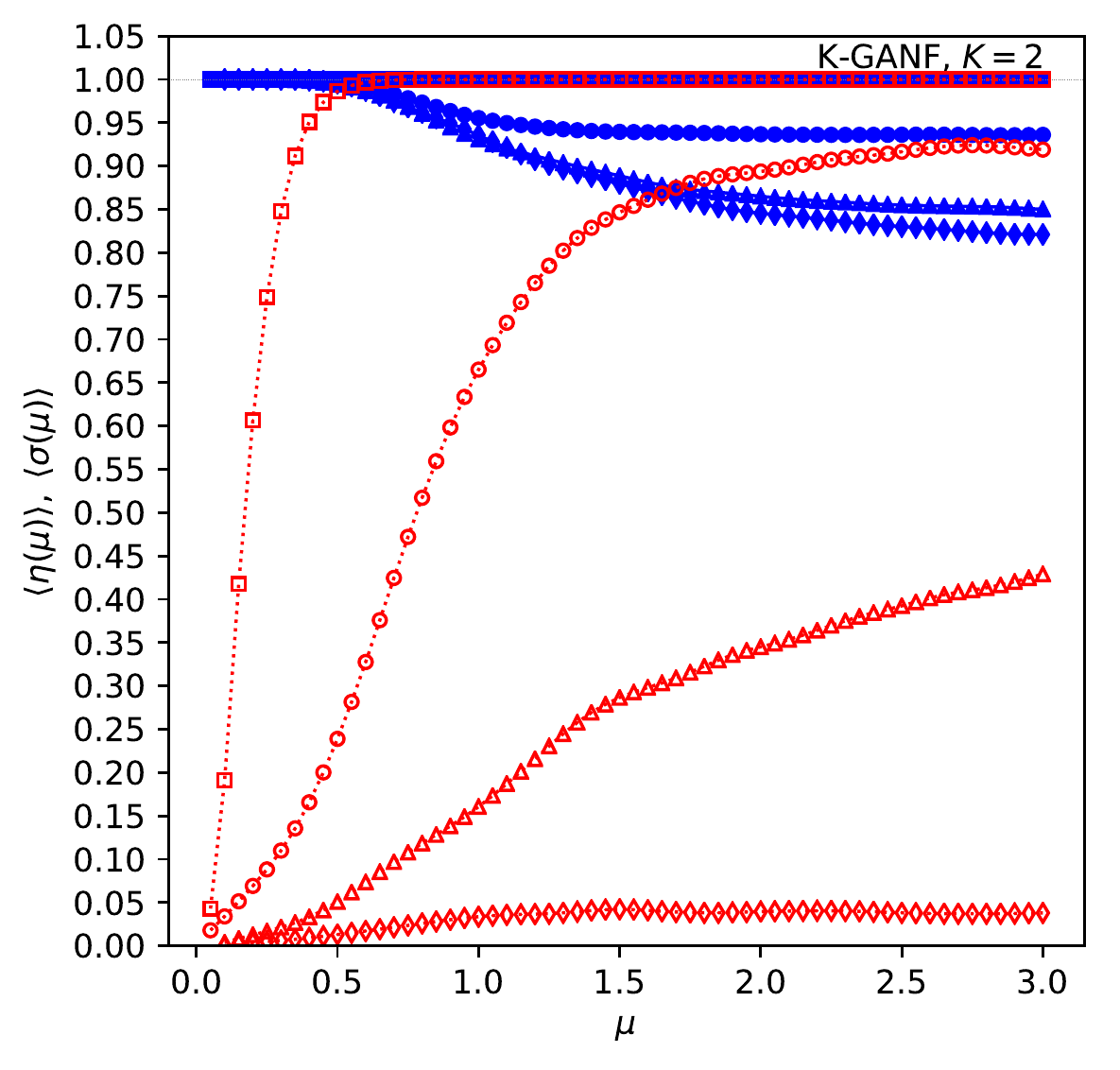}}\\[-2.5ex]
\caption{The probability  of exact synthesis $\langle \eta(\mu) \rangle$ (blue filled symbols) and 
the probability of exact recovery of unknown clauses $\langle \sigma(\mu) \rangle$ (red empty symbols) as a function of 
$\mu=M/N$ and the number of unknown clauses ($S=1$ squares, $S=2$ circles, $S=3$ triangles, $S=4$ diamonds), for $p=0.5$. 
K-DNF left column, K-GANF right column, $K=1$ top and $K=2$ bottom. 
}
\end{figure}

\section{Conclusion}

The problem of extracting logic inference rules from incomplete Boolean data observations is frequently 
encountered in the analysis of biological networks. 
The ill conditioning and incompleteness of the problem can also generate spurious inferences, 
which are a serious threat to valid inductive inference rules.  
Using random Boolean data as a null model, here we made an attempt to measure the competition between valid and spurious inductive inference rules from a given data set. 
More exactly, we have formulated two greedy search algorithms, which synthesize a given Boolean response variable in a sparse 
DNF, and respectively a sparse GANF of the variables from the observation data. 
Also, we have shown numerically that both algorithms can synthesize very sparse expansions with high probabilities, and 
the synthesis probability exhibits a transition as a 
function of the number of observations, with a transition point depending on the probability of the Bernoulli process used to generate the Boolean data. 
However, these greedy algorithms cannot recover correctly an a priori synthesized response, due to the inherent competition mechanism 
between the existing "equivalent" valid and spurious clauses.


\begin{thebibliography}{99}

\bibitem{key-1} S. Rathmanner, M. Hutter, 
				\textit{A Philosophical Treatise of Universal Induction}, 
				Entropy 13(6), 1076 (2011).
\bibitem{key-2} S. Keles, M.J. van der Laan, C. Vulpe, 
				\textit{Regulatory motif finding by logic regression}, 
				Bioinformatics 20(16), 2799 (2004).
\bibitem{key-3} Y. Tamada, S. Imoto, 
				\textit{Identifying Drug Active Pathways from Gene Networks Estimated by Gene Expression Data}, 
				Genome Informatics 16(1), 182 (2005).
\bibitem{key-4} J. Zhao, J. Li, M. Xiong, 
				\textit{Test for interaction between two unlinked loci}, 
				American Journal of Human Genetics 79(5), 831 (2006).
\bibitem{key-5} C. Kooperberg, I. Ruczinski,
				\textit{Identifying Interacting SNPs Using Monte Carlo Logic Regression},
				Genetic Epidemiology 28, 157 (2005).
\bibitem{key-6} S. Holger, K. Ickstadt, 
				\textit{Identification of SNP interactions using logic regression}, 
				Biostatistics 9(1), 187 (2008).
\bibitem{key-7} S. Mukherjee, S. Pelech, R.M. Neve, W.-L. Kuo, S. Ziyad, P.T. Spellman, J.W. Gray, T. P. Speed, 
				\textit{Sparse combinatorial inference with an application in cancer biology},
				Bioinformatics, 25(2), 265 (2009).
\bibitem{key-8} T.A. Knijnenburg, G.W. Klau, F. Iorio, M.J. Garnett, U. McDermott, I. Shmulevich, L.F.A. Wessels, 
				\textit{Logic models to predict continuous outputs based on binary inputs with an application to personalized cancer therapy}, 
				Scientific Reports 6, 36812 (2016).
\bibitem{key-9} N. Berlow, L. Davis, C. Keller, R. Pal, 
				\textit{Inference of dynamic biological networks based on responses to drug perturbations},
				EURASIP Journal on Bioinformatics and Systems Biology, 2014:14 (2014).
\bibitem{key-10} A.P. Kamath, N.K. Karmarkar, K. Ramakrishnan, M. G. C. Resende, 
				\textit{A continuous approach to inductive inference}, 
				Mathematical Programming 57, 215 (1992).
\bibitem{key-11} T. Sasao, 
				\textit{Representation of logic functions using EXOR operators}, 
				Proc. Workshop Applications of the Read-Muller Expansion in Circuit Design, Makuhari, Japan, pp. 308-313 (1995).
\bibitem{key-12} T. Sasao, D. Debnath, 
				\textit{Generalized Reed-Muller Expressions: Complexity and Exact Minimization Algorithm}, 
				IEICE Trans. Fundamentals, Vol. E79-A, No. 12, 2123 (1996).
\bibitem{key-13} P. Porwik, 
				\textit{Efficient calculation of the Reed-Muller form by means of the Walsh transform},
				Int. J. Appl. Math. Comput. Sci., Vol.12, No.4, 571 (2002).
\end{thebibliography}
\end{document}